\documentclass[12pt]{iopart}
\usepackage{iopams}
\usepackage{graphicx}
\usepackage{color}

\begin{document}

\title{Specific-heat exponent and modified hyperscaling in the 4D random-field Ising model}

\author{N.~G.~Fytas$^{1}$, V.~Mart\'{i}n-Mayor $^{2,3}$, M.~Picco$^{4}$ and N.~Sourlas$^{5}$}

\address{$^1$Applied Mathematics Research Centre,
Coventry University, Coventry CV1 5FB, United Kingdom}
\address{$^2$ Departamento de F\'{\i}sica Te\'orica I, Facultad de Ciencias
  F\'{\i}sicas, Universidad Complutense de Madrid, 28040 Madrid, Spain}
\address{$^3$ Instituto de Biocomputaci\'on y
  F\'{\i}sica de Sistemas Complejos (BIFI), 50018 Zaragoza, Spain}
\address{$^4$Sorbonne Universit\'es, Universit\'e Pierre et Marie Curie -- Paris VI, \\
Laboratoire de Physique Th\'eorique et Hautes Energies,\\
4 Place Jussieu, 
75252 Paris Cedex 05, France}
\address{$^5$ Laboratoire de Physique Th\'eorique de l'Ecole
Normale
  Sup\'erieure (Unit{\'e} Mixte de Recherche du CNRS et de l'Ecole Normale
  Sup\'erieure, associ\'ee \`a l'Universit\'e Pierre et Marie Curie, PARIS VI)
  24 rue Lhomond, 75231 Paris CEDEX 05, France}
\date{\today}

\begin{abstract}
We report a high-precision numerical estimation of the critical
exponent $\alpha$ of the specific heat of the random-field Ising
model in four dimensions. Our result $\alpha = 0.12(1)$ indicates
a diverging specific-heat behavior and is consistent with the
estimation coming from the modified hyperscaling relation using
our estimate of $\theta$ via the anomalous dimensions $\eta$ and
$\bar{\eta}$. Our analysis benefited form a high-statistics
zero-temperature numerical simulation of the model for two
distributions of the random fields, namely a Gaussian and
Poissonian distribution, as well as recent advances in finite-size
scaling and reweighting methods for disordered systems. An
original estimate of the critical slowing down exponent $z$ of the
maximum-flow algorithm used is also provided.
\end{abstract}

\submitto{Journal of Statistical Mechanics}

\maketitle

\section{Introduction}
\label{sec:Introduction}

The random-field Ising model (RFIM) is one of the archetypal
disordered
systems~\cite{imr75,aharony76,young77,fishman79,parisi79,cardy84,imbrie84,villain84,bray85,fisher86,schwartz85,gofman93,esser97,barber,young1999},
extensively studied due to its theoretical interest, as well as
its close connection to experiments in
hard~\cite{by1991,rieger1995,young1999,jaccarino} and soft
condensed matter systems~\cite{vink}. Its beauty is that the
mixture of random fields and the standard Ising model creates rich
physics and leaves many still unanswered problems. The existence
of an ordered ferromagnetic phase for the RFIM, at low temperature
and weak disorder, followed from the seminal paper of Imry and
Ma~\cite{imr75}, when the space dimension is greater than two ($D
> 2$)~\cite{villain84,bray85,fisher86,berker86,bricmont87}. This
has provided us with a general qualitative agreement on the sketch
of the phase boundary, separating the ordered ferromagnetic phase
from the high-temperature paramagnetic
one~\cite{newman93,machta00,newman96,itakura01,fytas08,aharony78}.

Although nowadays the view that the phase transition of the RFIM
is of second order, irrespective of the form of the random-field
distribution and for all values of the disorder
strength~\cite{fytasPRL}, there are still some puzzling behavior
that remains contradictory. One of the main problems, that we also
consider in the current work, refers to the scaling behavior of
the specific heat and the corresponding value of the critical
exponent $\alpha$; the latter having severe implications for the
scaling
relations~\cite{fytasPRL,fytasPRE,middleton1,hartmann01,theodorakis,fytasEPJB,nowak}.

The RFIM Hamiltonian is
\begin{equation}
\label{H} {\cal H} = - J \sum_{<xy>} S_x S_y - \sum_{x} h_x S_x \;
,
\end{equation}
with the spins $S_x = \pm 1$ occupying the nodes of a hyper-cubic
lattice in space dimension $D$ with nearest-neighbor ferromagnetic
interactions $J$ and $h_x$ independent random magnetic fields with
zero mean and dispersion $\sigma$. In the present work we consider
the Hamiltonian~(\ref{H}) on a $D=4$ hyper-cubic lattice with
periodic boundary conditions and energy units $J=1$. Our random
fields $h_{x}$ follow either a Gaussian $({\mathcal P}_G)$, or a
Poissonian $({\mathcal P}_P)$ distribution
\begin{equation}
\label{distribution} {\mathcal P}_G(h,\sigma) = {1\over \sqrt{2
\pi
      \sigma^2}} e^{- {h^2 \over 2\sigma^2}}\;;\ {\mathcal P}_P(h,\sigma) = {1\over 2 |\sigma| } e^{- {|h| \over
\sigma}}\;,
\end{equation}
where $-\infty< h < \infty$ and $\sigma$ our single
disorder-strength control parameter.

Note that the fact that we consider the model exactly at $T=0$ is
no restriction, because the temperature is an irrelevant
perturbation~\cite{young1999,villain84,bray85,fisher86,berker86}.
As it is well-established, in order to describe the critical
behavior of the model one needs two correlation functions, namely
the connected and disconnected propagators,
$C^{\mathrm{(con)}}_{xy}$ and $C^{\mathrm{(dis)}}_{xy}$. At the
critical point and for large $r$ ($r$: distance between $x$ and
$y$), they decay as
\begin{equation}\label{eq:anomalous}
C^{\mathrm{(con)}}_{xy}\!\equiv\!\frac{\partial\overline{\langle
  S_x\rangle}}{\partial h_y} \!\sim\!
\frac{1}{r^{D-2+\eta}}\,;\
 C^{\mathrm{(dis)}}_{xy}\!\equiv\!
\overline{\langle S_x\rangle\langle S_y\rangle}\! \sim\!
\frac{1}{r^{D-4+\bar\eta}}\,,
\end{equation}
where the $\langle \ldots \rangle$ are thermal mean values as
computed for a given realization, a sample, of the random fields
$\{h_x\}$. Over-line refers to the average over the samples. The
correlation length corresponding to $C^{\mathrm{(con)}}$ is
denoted by $\xi^{\mathrm{(con)}}$ (we use $\xi^{\mathrm{(dis)}}$
for $C^{\mathrm{(dis)}}$, respectively).

The relationship between the anomalous dimensions $\eta$ and
$\bar\eta$ has been hotly debated in the literature, especially
with respect to the so-called two-exponent scaling scenario
$\bar\eta = 2\eta$~\cite{schwartz85,gofman93} and its implications
on the violation of the hyperscaling exponent $\theta$ via
\begin{equation}\label{eq:theta}
\theta = 2 -\bar\eta+\eta = 2-\eta+\Delta_{\eta,\bar\eta},
\end{equation}
where $\Delta_{\eta,\bar\eta} = 2\eta-\bar\eta$. In fact, the
original prediction $\Delta_{\eta,\bar\eta} = 0$ by Schwartz and
coworkers~\cite{schwartz85,gofman93} has been recently questioned
by Tarjus and coworkers~\cite{tissier11,tissier12,tarjus13}. These
latter authors suggested that rare events, neglected in
Refs.~\cite{schwartz85,gofman93}, spontaneously break
supersymmetry at the intermediate dimension
$D_{\mathrm{int}}\approx 5.1$. For $D>D_{\mathrm{int}}$ replica
predictions~\cite{parisi79} hold: supersymmetry is valid and $
\bar{\eta} = \eta $. For $ D < D_{\mathrm{int}} $, instead, there
are three independent critical exponents. Recent high-precision
numerical simulations by the current authors~\cite{fytas4D} provided clear cut
evidence that $\Delta_{\eta,\bar\eta} >0 $ in favor of the
three-exponent scaling scenario and the spontaneous supersymmetry
breaking~\cite{tissier11,tissier12} at some $D_{\mathrm{int}}
> 4$. Some of the results of
Ref.~\cite{fytas4D} will be used to corroborate the current
analysis.

The rest of the paper is laid out as follows: In the next
Section~\ref{sec:FSS} we describe the finite-size scaling methods
used and in Section~\ref{sec:numerical} we briefly outline the
numerics performed. In Section~\ref{sec:results} we present our
results for: (i) the critical exponent $\alpha$ of the specific
heat, Section~\ref{sec:alpha}, modified hyperscaling,
Section~\ref{sec:hyper}, and the dynamical aspects of the
algorithm used in Section~\ref{sec:z}. Section~\ref{sec:conclusions}
concludes the article.

\section{Finite-size scaling}
\label{sec:FSS}

In the present work we are mostly interested in investigating the
controversial issue of the specific heat of the RFIM. The specific
heat of the RFIM can be experimentally measured~\cite{jaccarino}
and is of great theoretical importance. Yet, it is well known that
it is one of the most intricate thermodynamic quantities to deal
with in numerical simulations, even when it comes to pure systems.
For the RFIM, Monte Carlo methods at $T>0$ have been used to
estimate the value of its critical exponent $\alpha$, but were
restricted to rather small systems sizes and have also revealed
many serious problems, i.e., severe violations of self
averaging~\cite{PS02,fytas06}. A better picture emerged throughout
the years from $T=0$ computations, proposing estimates of
$\alpha\approx 0$~\cite{fytasPRL,middleton1}. However, even by
using the same numerical techniques, but different scaling
approaches, some inconsistencies have been recorded in the
literature~\cite{hartmann01,middleton1,theodorakis,fytasEPJB}. The
origin of these inconsistencies~\cite{hartmann01,picco13} can be
traced back to the fact that the specific heat contains, according
to simulations, a regular term as well as a singular term with a
large exponent of the order of $-0.6$. This can be interpreted as
a specific-heat exponent with a negative value $\alpha/\nu\simeq
-0.6$. Another possibility is a very small value of the
specific-heat exponent (which then gives a near constant term) and
a corrections-to-scaling term $L^{-\omega}$. Indeed, in three
dimensions, $\omega$ has a value close to $0.6$~\cite{fytasPRL}.
In order to distinguish between these two scenarios, one needs to
use a hyperscaling relation in order to get another direct
determination of $\alpha$ as a function of the magnetic critical
exponents that were measured directly in our simulations. For the
case of the 3D RFIM, it was then shown that $\alpha \simeq
0$~\cite{hartmann01,picco13}. For the 3D RFIM, there exists also
experiments on diluted antiferromagnetic systems, which are
expected to be in the same universality class as the random-field
magnets~\cite{fishman79,cardy84}, and which suggested a
logarithmic divergence of the specific heat~\cite{jaccarino}. For
the 4D RFIM that we consider in the present study, we will also
compare our direct measurements with the hyperscaling relation to
check the validity of our results.

The specific heat can be also estimated using ground-state
calculations and applying thermodynamic relations employed by
Hartmann and Young~\cite{hartmann01} and Middleton and
Fisher~\cite{middleton1}. The method relies on studying the
singularities in the bond-energy density $E_{J}$~\cite{holm97}.
This bond energy density is the first derivative $\partial
E/\partial J$ of the ground-state energy with respect to the
random-field strength, say $\sigma$~\cite{middleton1,hartmann01}.
The derivative of the sample averaged quantity $\overline{E}_{J}$
with respect to $\sigma$ then gives the second derivative with
respect to $\sigma$ of the total energy and thus the
sample-averaged specific heat $C$. The singularities in $C$ can
also be studied by computing the singular part of
$\overline{E}_{J}$, as $\overline{E}_{J}$ is just the integral of
$C$ with respect to $\sigma$. The general finite-size scaling form
assumed is that the singular part of the specific heat behaves as
\begin{equation}
\label{eq:C_scaling} C_{\rm s}\sim L^{\alpha/\nu}\tilde{C}\left
[(\sigma-\sigma_{\rm c})L^{1/\nu}\right].
\end{equation}
Thus, one may estimate $\alpha$ by studying the behavior of
$\overline{E}_{J}$ at $\sigma = \sigma_{\rm c}$~\cite{middleton1}.
The computation from the behavior of $\overline{E}_{J}$ is based
on integrating the above scaling equation up to $\sigma_{\rm c}$ ,
which gives a dependence of the form
\begin{equation}
\label{eq:E_J_scaling} \overline{E}_{J}(L,\sigma_{\rm c}) = A + B
L^{(\alpha-1)/\nu},
\end{equation}
with $A$ and $B$ non universal constants.

\begin{figure}[h]
\begin{center}
\includegraphics[height=0.6\linewidth,angle=0]{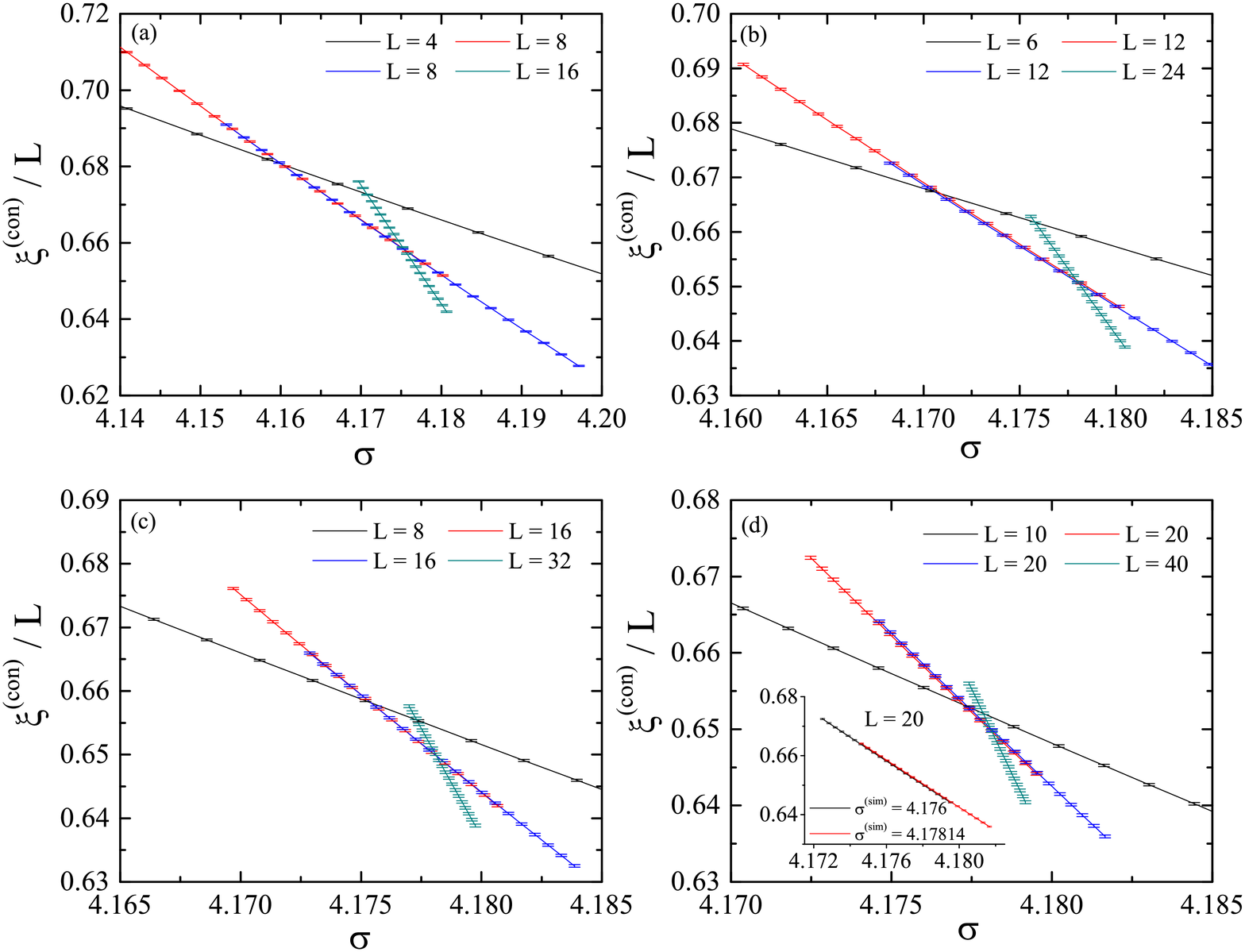}
\end{center}
\caption{Connected correlation length in units of the system size
$L$ versus $\sigma$ for the case of the Gaussian RFIM and for the
sequence of lattice sizes used in the application of the modified
quotients method in equations~(\ref{eq:FSS-3L}) and
(\ref{eq:QO_new}). Data taken from Ref.~\cite{fytas4D}. The inset
of panel (d) shows $\xi^{\rm (con)}/L$ for $L = 20$ and two
distinct simulations corresponding to different simulation values,
$\sigma^{\rm (sim)}$, and different sets of $10^7$ random-field
realizations.
 \label{fig:1}}
\end{figure}

Applying standard finite-size scaling methods via
equation~(\ref{eq:E_J_scaling}) would require an a priori
knowledge of the ``exact'' value of the critical field strength
$\sigma_{\rm c}$. Since what we always have at hand is a numerical
estimate this approach would then incorporate a degree of
approximation coming from the uncertainty in the value of
$\sigma_{\rm c}$. To overcome these scaling problems, we
implemented here a variant~\cite{fytasPRE} of the original
quotients method, also known as phenomenological
renormalization~\cite{ballesteros96,amit05,nightingale76}. This
method allows for a particularly transparent study of corrections
to scaling, that up to now have been considered as the Achilles'
heel in the study of the $D\geq 3$ random-field
problem~\cite{fytasPRL}. The general idea is to compare
observables computed in pair of lattices $(L,2L)$. We start
imposing scale-invariance by seeking the $L$-dependent critical
point: the value of $\sigma$ such that $\xi_{2L}/\xi_L=2$, i.e.,
the crossing point for $\xi_L/L$ (see also figure~\ref{fig:1}).
Now, for dimensionful quantities $O$, scaling in the thermodynamic
limit as $\xi^{x_O/\nu}$, we consider the quotient
$Q_O=O_{2L}/O_L$ at the crossing. Thus, we have:
\begin{equation}\label{eq:QO}
Q_O^\mathrm{(cross)}=2^{x_O/\nu}+\mathcal{O}(L^{-\omega})\,,\
\end{equation}
where $x_O/\nu$, and the scaling-corrections exponent $\omega$ are
universal.

Since $\alpha-1$ is negative, equation~(\ref{eq:E_J_scaling}) is
dominated by the non-divergent back ground $A$, forcing us to
modify the standard phenomenological renormalization. We get rid
of $A$ by considering three lattice sizes in the following
sequence: $(L_1, L_2, L_3) = (L, 2L, 4L)$ (see figure~\ref{fig:1}
for an instructive illustration of the three-lattice variant of
the quotients method based on the crossings of the $\xi^{\rm
(con)} / L$). We generalize equation~(\ref{eq:QO}) by taking now
the quotient of the differences at the crossings of the pairs
$(L,2L)$ and $(2L,4L)$, respectively
\begin{equation}\label{eq:FSS-3L}
\hat Q_O=\frac{\left.(O_{4L}-O_{2L})\right|_{(\xi_{4L}/\xi_{2L})=2}}{\left(O_{2L} - O_{L})\right|_{(\xi_{2L}/\xi_{L})=2}}\,.
\end{equation}
Applying this formula to the bond energy we obtain
\begin{equation}\label{eq:QO_new}
\hat Q_{\overline{E}_J}^\mathrm{(cross)}=2^{(\alpha -
1)/\nu}+\mathcal{O}(L^{-\omega}).
\end{equation}

\section{Numerical simulations}
\label{sec:numerical}

Our numerical simulations were described in Ref.~\cite{fytas4D}.
We therefore only recall here the crucial details. In order to
apply the above formula~(\ref{eq:QO_new}), we simulated systems
with linear sizes up to $L = 40$, which provided us with four sets
of three-lattice size sequences, as shown in figure~\ref{fig:1}(a)
- (d). For each set ($L, \sigma$) and for each field distribution,
Gaussian and Poissonian, we simulated $10^7$ independent
random-field realizations (exceeding previous relevant
studies~\cite{middleton4D,hartmann4D} by a factor of $10^3$ at
least). Suitable generalized fluctuation-dissipation formulas and
reweighting extrapolations have been applied and facilitated our
analysis, as exemplified in Ref.~\cite{fytasPRE}. A comparative
illustration with respect to the errors induced by the reweighting
method and the disorder averaging process is shown in the inset of
panel (d) of figure~\ref{fig:1} for the universal ratio $\xi^{\rm
(con)} / L$ of an L = 20 Gaussian RFIM and for two sets of
simulations, as outlined in the figure. Clearly, this latter
accuracy test serves in favor of the proposed scheme. The
calculation of the ground states of the RFIM was based on the
well-established
mapping~\cite{ogielski85,hartmann95,bastea98,hartmann99,fytasPRL,middleton1,hartmann01,theodorakis,fytasEPJB,hartmann02,seppala,middleton2,middleton3,machta03,alava,zumsande,puri11,ahrens,weigel,fytas12,hartmannbook1,hartmannbook2}
to the maximum-flow problem~\cite{angles,cormen,papadimitriou}.
This is a combinatorial optimization problem which can be solved
exactly using efficient, i.e., polynomial-time, algorithms. The
most efficient network flow algorithm used to solve the RFIM is
the push-relabel algorithm of Tarjan and Goldberg~\cite{tarjan}.
We prepared our own C version of the algorithm, involving a
modification that removes the source and sink nodes, reducing
memory usage and also clarifying the physical
connection~\cite{middleton2,middleton3}. Additionally, the
computational efficiency of our algorithm has been increased via
the use of periodic global updates~\cite{middleton2,middleton3}.

\section{Results}
\label{sec:results}

\subsection{Specific-heat exponent}
\label{sec:alpha}

\begin{table}
\caption{\label{tab:specheat} Effective critical exponent ratio
  $(\alpha-1)/\nu$ using a three lattice-size variant
  $(L_{1},L_{2},L_{3})=(L,2L,4L)$, see equation~(\ref{eq:QO_new}), of the original quotients method.}
\begin{tabular}{llc}
\hline \hline
Crossing Point & $(L_{1},L_{2},L_{3})$ & $(\alpha-1)/\nu$\\
\hline G$^{\rm (con)}$   &$(4,8,16)$    &   -0.697(6)          \\
     &$(6,12,24)$    &   -0.848(6)         \\
     &$(8,16,32)$    &   -0.912(8)         \\
     &$(10,20,40)$   &   -0.941(12)         \\
\hline
G$^{\rm (dis)}$   &$(4,8,16)$    &   -0.954(4)          \\
     &$(6,12,24)$    &   -0.975(4)         \\
     &$(8,16,32)$    &   -0.990(5)         \\
     &$(10,20,40)$   &   -0.998(4)         \\
\hline
P$^{\rm (con)}$   &$(4,8,16)$    &   -1.031(7)          \\
     &$(6,12,24)$    &   -1.014(9)         \\
     &$(8,16,32)$    &   -1.002(11)         \\
     &$(10,20,40)$   &   -0.998(8)         \\
\hline
P$^{\rm (dis)}$   &$(4,8,16)$    &   -1.205(4)          \\
     &$(6,12,24)$    &   -1.129(3)         \\
     &$(8,16,32)$    &   -1.084(6)         \\
     &$(10,20,40)$   &   -1.057(12)         \\
     \hline
\end{tabular}
\end{table}

As we applied the quotients method at both the crossings of the
connected and disconnected correlation length over the system
size, i.e., $\xi^{\rm (con)} / L$ and $\xi^{\rm (dis)} / L$,
typically the sets of simulations were doubled for each system
size as the crossings between the connected and disconnected cases
varied. Note also, that throughout the main manuscript we have
used the notation $\rm {Z}^{\rm (x)}$, where Z denotes the
distribution, i.e., G for Gaussian and P for Poissonian, and the
superscript x refers to the connected (con) and disconnected (dis)
type of the universal ratio $\xi^{\rm (x)} / L$.

\begin{figure}[h]
\begin{center}
\includegraphics[height=0.6\linewidth,angle=0]{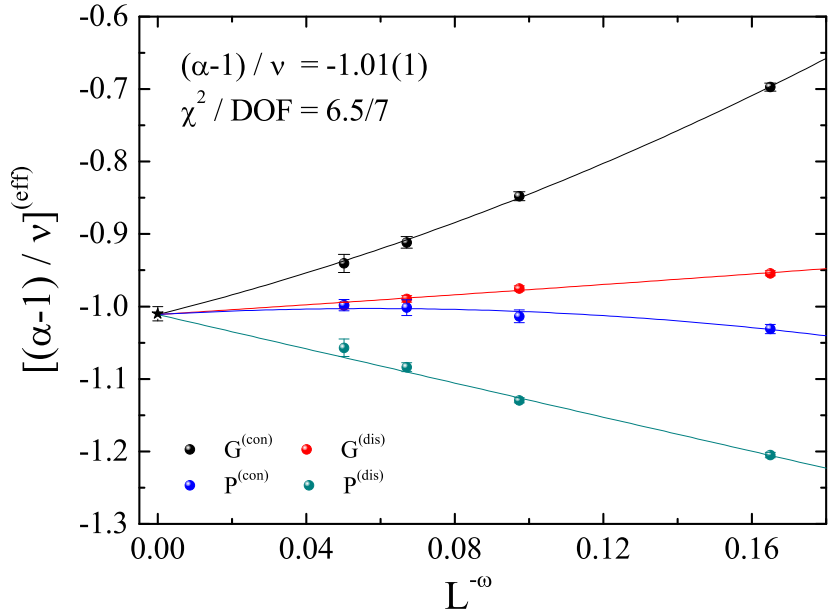}
\end{center}
\caption{Infinite limit-size extrapolation of the effective
exponent ratio $(\alpha - 1)/\nu$.
 \label{fig:2}}
\end{figure}

Our results for the effective exponent ratio $(\alpha-1)/\nu$ are
given in Table~\ref{tab:specheat} and their extrapolation is shown
in figure~\ref{fig:2}. The solid lines in figure~\ref{fig:2} show
a joint polynomial fit, second order in $L^{-\omega}$, where
$\omega$ was set to the estimated value $\omega =
1.30$~\cite{fytas4D}. The extrapolated value for the exponent
ratio is $(\alpha - 1)/\nu = -1.01(1)$ and is marked by the filled
star at $L^{-\omega} = 0$. The quality of the fit is excellent,
$\chi^2/ {\rm DOF} = 6.5/7$, where DOF denoted the degrees of
freedom in the fit, and is also given in the plot. Using now our
previous estimate $\nu = 0.8718(58)$ for the critical exponent of
the correlation length~\cite{fytas4D}, simple algebra (and error
propagation) gives the value
\begin{equation}
\label{eq:alpha} \alpha = 0.12(1),
\end{equation}
for the critical exponent of the specific heat.

Middleton~\cite{middleton4D}, using also ground-state simulations
and the scaling of the bond energy at the candidate critical field
value estimated in his analysis, $\sigma_{\rm c} = 4.179$,
proposed a value of $\alpha=0.26(5)$. As it will be shown below,
this value is not in agreement with modified hyperscaling.
Noteworthy, another relevant ground-state numerical work of the 4D
RFIM by Hartmann~\cite{hartmann4D} was unable to conclude whether
the numerical data are better described by a logarithmic
divergence or by an algebraic behavior with a small exponent.

\subsection{Modified hyperscaling}
\label{sec:hyper}

A crucial point in the scaling theory of the RFIM is the
hyperscaling violation exponent $\theta$ [recall
equation~(\ref{eq:theta})] and its relation to the critical
exponent $\alpha$ of the specific heat via the modified
hyperscaling relation
\begin{equation}
\label{eq:hyperscaling} (D - \theta)\nu = 2 - \alpha.
\end{equation}
Numerous works have tried in the past to reconcile the numerical
estimates of $\alpha$ with those stemming from the above
equation~(\ref{eq:hyperscaling}) but have mostly
failed~\cite{rieger1995,hartmann01,machta03,fytas06}.

We now compare our numerical estimate $\alpha=0.12(1)$ with the
one obtained via equations~(\ref{eq:theta}) and
(\ref{eq:hyperscaling}) and our previous estimates for the
anomalous dimension $\eta=0.1930(13)$ and the two-exponent
difference $\Delta_{\eta,\bar\eta} = 2\eta - \bar\eta =
0.0322(23)$~\cite{fytas4D}. In particular, plugging these values
to equation~(\ref{eq:theta}) we obtain
\begin{equation}
\label{eq:theta_estimate} \theta = 1.839(3),
\end{equation}
which is compatible to the value $\theta=1.82(7)$ of
Middleton~\cite{middleton4D}, but more accurate. On a second step,
we can use this value of $\theta$ and $\nu$ to get a further
estimate for the critical exponent of the specific heat:
\begin{equation}
\alpha=2-\nu(D- 2 +\eta -\Delta_{\eta,\bar{\eta}})\,.
\end{equation}
The error propagation for the above equation is rather simple (because
the final error is basically dominated by the uncertainty in
$\nu$). We finally obtain $\alpha = 0.12(1)$, in excellent agreement
to the numerical estimate based on the finite-size scaling analysis
described above.

\begin{figure}
\begin{center}
\includegraphics[height=0.6\linewidth,angle=0]{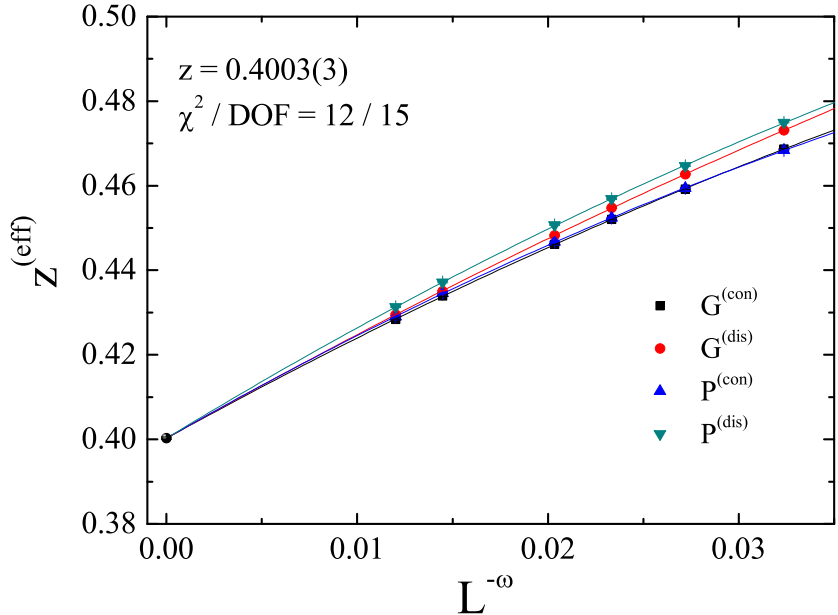}
\end{center}
\caption{Infinite limit-size extrapolation of the effective
exponent $z$ of the push-relabel algorithm.
 \label{fig:3}}
\end{figure}

\subsection{Critical dynamic slowing down}
\label{sec:z}

Finally, we present some computational aspects of the implemented
push-relabel algorithm and its performance on the study of the
RFIM. Although its generic implementation has a polynomial time
bound, its actual performance depends on the order in which
operations are performed and which heuristics are used to maintain
auxiliary fields for the algorithm. Even within this polynomial
time bound, there is a power-law critical slowing down of the
push-relabel algorithm at the zero-temperature
transition~\cite{ogielski85,middleton2}. A direct way to measure
the dynamics of the algorithm is to examine the dependence of the
running time, measured by the number of push-relabel operations,
on system size $L$~\cite{middleton2,middleton3}. Such an analysis
has already been performed for the 3D version of the
model~\cite{fytasPRE} and a FIFO queue implementation. We present
here results for the 4D version of the model using our scaling
approach within the quotients method and numerical data for both
Gaussian and Poissonian random-field distributions for system
sizes up to $L=60$. In figure~\ref{fig:3} we plot the effective
exponent values of $z$ at the various crossing points considered,
as indicated. The solid line is a joint, second-order in
$L^{-\omega}$, polynomial fit for system sizes $L\ge 14$ with
$\chi^2/{\rm DOF} = 12/15$. The obtained estimate for the dynamic
critical exponent $z$ is $0.4003(3)$, as marked by the filled
circle at $L^{-\omega} = 0$.

\section{Conclusions}
\label{sec:conclusions}

Using extensive numerical simulations at zero temperature and
efficient finite-size scaling methods we presented a
high-precision numerical estimate of the critical exponent
$\alpha$ of the specific heat of the random-field Ising model in
four dimensions. Our result is fully consistent with the
estimation coming from the modified hyperscaling relation, giving
us full credit on the numerical and scaling approach implemented.
Finally, we provided an original estimate of the critical slowing
down exponent of the maximum-flow algorithm used.

\section*{Acknowledgments}
Our $L=52, 60$ lattices were simulated in the {\em MareNostrum}
and {\em Picasso} supercomputers (we thankfully acknowledge the
computer resources and assistance provided by the staff at the
{\em Red Espa\~nola de Supercomputaci\'on\/}). N.~G.~F. is
grateful to Coventry University for providing a Research
Sabbatical Fellowship during which this work has been completed.
V.~M.-M. was partly supported by MINECO (Spain) through Grant No.
FIS2015-65078-C2-1-P.

\end{document}